\documentclass[aps,prl,twocolumn,export,superscriptaddress]{revtex4-2} 
\usepackage[colorlinks=true,linkcolor=blue,citecolor=blue,urlcolor=blue]{hyperref}

\usepackage[utf8]{inputenc}
\usepackage[german, english]{babel}
\usepackage[dvipsnames]{xcolor}
\usepackage{amsthm}
\usepackage{mathtools}
\usepackage{tikz}
\usetikzlibrary{decorations.markings} 
\usetikzlibrary{decorations.shapes,shapes.geometric} 

\usepackage{amsfonts}
\usepackage{amssymb}
\usepackage{braket}
\usepackage{subfigure}
\usepackage[export]{adjustbox} %valign=c
\usepackage{dsfont}
\usepackage[inline]{enumitem}  % inline lists
\usepackage{times}
\usepackage[smallerops]{newtxmath}
\usepackage{microtype}

\usepackage[normalem]{ulem}

\DeclareMathAlphabet{\mathcal}{OMS}{cmsy}{m}{n}
\DeclareMathAlphabet{\mathsf}{OT1}{cmss}{m}{n}
\DeclareMathAlphabet{\mathbb}{U}{msb}{m}{n}

\DeclareMathOperator{\sech}{sech}

\usetikzlibrary{decorations.pathreplacing,decorations.markings,arrows,decorations.pathmorphing}

\tikzset{%
    x=2.4ex,
    y=2.4ex,
    baseline={(0,-0.1)}
}

\definecolor{xpauli}{HTML}{D62728}
\definecolor{zpauli}{HTML}{1F77B4}

\newcommand{\faceop}[2]{%
    \draw [line width=1.25, xpauli, preaction={draw, line width=4, white}] ({-1+#1}, {-1+#2}) rectangle ({1+#1}, {1+#2});
}

\newcommand{\XL}[2]{%
    \begin{scope}[very thick,decoration={markings,
    mark=at position 0.16667 with {\fill[white] +(-2pt,-2pt) rectangle +(2pt,2pt);},
    mark=at position 0.5 with {\fill[white] +(-2pt,-2pt) rectangle +(2pt,2pt);},
    mark=at position 0.83333 with {\fill[white] +(-2pt,-2pt) rectangle +(2pt,2pt);}}
    ]
        \draw [line width=1.25, xpauli, preaction={decorate}] ({1+#1}, {-1+#2}) to ({-1+#1}, {-1+#2}) to ({-1+#1}, {1+#2}) to ({1+#1}, {1+#2});
    \end{scope}
}

\newcommand{\XR}[2]{%
    \begin{scope}[very thick,decoration={markings,
    mark=at position 0.5 with {\fill[white] +(-2pt,-2pt) rectangle +(2pt,2pt);}}
    ]
        \draw [line width=1.25, xpauli, preaction={decorate}] ({1+#1}, {-1+#2}) to ({1+#1}, {1+#2});
    \end{scope}
}

\newcommand{\ZL}[2]{%
    \begin{scope}[very thick,decoration={markings,
    mark=at position 0.5 with {\fill[white] +(-2pt,-2pt) rectangle +(2pt,2pt);}}
    ]
        \draw [line width=1.25, zpauli, preaction={decorate}] ({-1+#1}, {-1+#2}) to ({-1+#1}, {1+#2});
    \end{scope}
}

\newcommand{\ZR}[2]{%
    \begin{scope}[very thick,decoration={markings,
    mark=at position 0.16667 with {\fill[white] +(-2pt,-2pt) rectangle +(2pt,2pt);},
    mark=at position 0.5 with {\fill[white] +(-2pt,-2pt) rectangle +(2pt,2pt);},
    mark=at position 0.83333 with {\fill[white] +(-2pt,-2pt) rectangle +(2pt,2pt);}}
    ]
        \draw [line width=1.25, zpauli, preaction={decorate}] ({-1+#1}, {-1+#2}) to ({1+#1}, {-1+#2}) to ({1+#1}, {1+#2}) to ({-1+#1}, {1+#2});
    \end{scope}
}

\newcommand{\Xbackground}[2]{%
    \draw [line width=1., lightgray, opacity=0.5] ({1+#1}, {-1+#2}) to ({-1+#1}, {-1+#2}) to ({-1+#1}, {1+#2}) to ({1+#1}, {1+#2}) -- cycle;
}

\newcommand{\Xtop}[2]{%
    \begin{scope}[very thick,decoration={markings,
    mark=at position 0.5 with {\fill[white] +(-2pt,-2pt) rectangle +(2pt,2pt);}}
    ]
        \draw [line width=1.25, xpauli, preaction={decorate}] ({-1+#1}, {1+#2}) to ({1+#1}, {1+#2});
    \end{scope}
}

\newcommand{\Xbot}[2]{%
    \begin{scope}[very thick,decoration={markings,
    mark=at position 0.5 with {\fill[white] +(-2pt,-2pt) rectangle +(2pt,2pt);}}
    ]
        \draw [line width=1.25, xpauli, preaction={decorate}] ({-1+#1}, {-1+#2}) to ({1+#1}, {-1+#2});
    \end{scope}
}

\newcommand{\Xsides}[2]{%
    \begin{scope}[very thick,decoration={markings,
    mark=at position 0.5 with {\fill[white] +(-2pt,-2pt) rectangle +(2pt,2pt);}}
    ]
        \draw [line width=1.25, xpauli, preaction={decorate}] ({-1+#1}, {-1+#2}) to ({-1+#1}, {1+#2});
    \end{scope}
    \begin{scope}[very thick,decoration={markings,
    mark=at position 0.5 with {\fill[white] +(-2pt,-2pt) rectangle +(2pt,2pt);}}
    ]
        \draw [line width=1.25, xpauli, preaction={decorate}] ({1+#1}, {-1+#2}) to ({1+#1}, {1+#2});
    \end{scope}
}

\newcommand{\Zbackground}[2]{%
    \draw [line width=1., lightgray, opacity=0.5] ({#1}, {2+#2}) to ({#1}, {-2+#2});
    \draw [line width=1., lightgray, opacity=0.5] ({-1+#1}, {1+#2}) to ({1+#1}, {1+#2});
    \draw [line width=1., lightgray, opacity=0.5] ({-1+#1}, {-1+#2}) to ({1+#1}, {-1+#2});
}

\newcommand{\Ztop}[2]{%
    \begin{scope}[very thick,decoration={markings,
    mark=at position 0.16667 with {\fill[white] +(-2pt,-2pt) rectangle +(2pt,2pt);},
    mark=at position 0.5 with {\fill[white] +(-2pt,-2pt) rectangle +(2pt,2pt);},
    mark=at position 0.83333 with {\fill[white] +(-2pt,-2pt) rectangle +(2pt,2pt);}}
    ]
        \draw [line width=1.25, zpauli, preaction={decorate}] ({-1+#1}, {-1+#2}) to ({-1+#1}, {1+#2}) to ({1+#1}, {1+#2}) to ({1+#1}, {-1+#2});
    \end{scope}
}

\newcommand{\Zbot}[2]{%
    \begin{scope}[very thick,decoration={markings,
    mark=at position 0.16667 with {\fill[white] +(-2pt,-2pt) rectangle +(2pt,2pt);},
    mark=at position 0.5 with {\fill[white] +(-2pt,-2pt) rectangle +(2pt,2pt);},
    mark=at position 0.83333 with {\fill[white] +(-2pt,-2pt) rectangle +(2pt,2pt);}}
    ]
        \draw [line width=1.25, zpauli, preaction={decorate}] ({-1+#1}, {1+#2}) to ({-1+#1}, {-1+#2}) to ({1+#1}, {-1+#2}) to ({1+#1}, {1+#2});
    \end{scope}
}

\newcommand{\ZH}[2]{%
    \begin{scope}[very thick,decoration={markings,
    mark=at position 0.5 with {\fill[white] +(-2pt,-2pt) rectangle +(2pt,2pt);}}
    ]
        \draw [line width=1.25, zpauli, preaction={decorate}] ({-1+#1}, {#2}) to ({1+#1}, {#2});
    \end{scope}
}

\renewcommand{\vec}[1]{\mathbf{#1}}

\DeclarePairedDelimiter\abs{\lvert}{\rvert}%
\DeclarePairedDelimiter\norm{\lVert}{\rVert}%

\makeatletter
\let\oldabs\abs
\def\abs{\@ifstar{\oldabs}{\oldabs*}}
\let\oldnorm\norm
\def\norm{\@ifstar{\oldnorm}{\oldnorm*}}
\makeatother

\usepackage{bbold}

\begin{document}

\title{Playing nonlocal games across a topological phase transition on a quantum computer}

\date{March 7, 2024}

\author{Oliver Hart}
\affiliation{Department of Physics and Center for Theory of Quantum Matter, University of Colorado Boulder, Boulder, Colorado 80309 USA}

\author{David T. Stephen}
\affiliation{Department of Physics and Center for Theory of Quantum Matter, University of Colorado Boulder, Boulder, Colorado 80309 USA}
\affiliation{Department of Physics, California Institute of Technology, Pasadena, California 91125, USA}

\author{Dominic J. Williamson}
\thanks{Current address: IBM Quantum, IBM Almaden Research Center, San Jose, CA 95120, USA}
\affiliation{Centre for Engineered Quantum Systems, School of Physics, University of Sydney, Sydney, New South Wales 2006, Australia}

\author{Michael Foss-Feig}
\affiliation{Quantinuum, 303 S Technology Ct, Broomfield, CO 80021, USA}

\author{Rahul Nandkishore}
\affiliation{Department of Physics and Center for Theory of Quantum Matter, University of Colorado Boulder, Boulder, Colorado 80309 USA}

\begin{abstract}
Many-body quantum games provide a natural perspective on phases of matter in quantum hardware, crisply relating the quantum correlations inherent in phases of matter to the securing of quantum advantage at a device-oriented task. In this paper we introduce a family of multiplayer quantum games for which topologically ordered phases of matter are a resource yielding quantum advantage. Unlike previous examples, quantum advantage persists away from the exactly solvable point and is robust to arbitrary local perturbations, irrespective of system size. We demonstrate this robustness experimentally on Quantinuum’s H1-1 quantum computer by playing the game with a continuous family of randomly deformed toric code states that can be created with constant-depth circuits leveraging mid-circuit measurements and unitary feedback. We are thus able to tune through a topological phase transition -- witnessed by the loss of robust quantum advantage -- on currently available quantum hardware. This behavior is contrasted with an analogous family of deformed GHZ states, for which arbitrarily weak local perturbations destroy quantum advantage in the thermodynamic limit. Finally, we discuss a topological interpretation of the game, which leads to a natural generalization involving an arbitrary number of players.
\end{abstract}

\maketitle

%%%%%%%%%%%%%%%%%%%%%%%%%%%%%%%%%%%%%%%%%%%%%%%%%%%%%%%%%%%%%%%
%%%%%%%%%%%%%%%%%%%%%%%%%%%%%%%%%%%%%%%%%%%%%%%%%%%%%%%%%%%%%%%
%%%%%%%%%%%%%%%%%%%%%%%%%%%%%%%%%%%%%%%%%%%%%%%%%%%%%%%%%%%%%%%

The advent of noisy intermediate-scale quantum (NISQ) devices has opened up a new frontier for many-body physics, replete with fundamental questions that remain to be fully answered~\cite{PreskillNISQ, IppolitiNISQ}. For instance, can the notion of phases of matter (the foundation of condensed matter physics) be meaningfully applied to such devices? In other words, can phases of matter be realized and identified in a robust manner on quantum hardware, and can available hardware be continuously tuned between distinct phases of matter? Does the notion of phases of matter have any bearing on the computational power of the device in question? For example, can the correlations inherent in phases of matter be harnessed to gain quantum advantage at certain tasks? How robust are these notions to gate imperfections or to noise? Can we harness the unique features of NISQ devices -- such as direct access to the many-body wavefunction -- to design novel probes of exotic physics? These are some of the most pressing open problems at the intersection of condensed matter physics and quantum information science, and definitive answers are only beginning to emerge.

One stimulating recent development examines NISQ devices and many-body quantum states through the lens of \emph{multiplayer nonlocal quantum games}~\cite{DanielStringOrder, Daniel2022Exp, BBSgame, BulchandaniGames}. These are generalizations of Bell tests~\cite{Mermin, MerminPolynomials, brassard2005pseudotelepathy, brassard2005recasting} to the many-body context, which take advantage of the fact that measurements made on an entangled resource state can be more correlated than any classical local hidden-variable model would permit -- a property known as {\it contextuality}. Contextuality can be applied to gain quantum advantage at a suitably designed task, whereby a set of ``players'' who are not allowed to communicate with each other are required to output appropriately correlated answers to a set of questions, such that there is no classical strategy that guarantees success at this task. Armed with an appropriately contextual resource state, the players can ``win'' the game with better-than-classical probability. Insofar as the pattern of entanglement (and hence contextuality) is a property of a quantum phase of matter, there can therefore be a tight link between phases of matter and the ability to secure quantum advantage at a given task. This notion is crystallized in the context of measurement-based quantum computation (MBQC)~\cite{RaussendorfMBQC}, where the success probability for a given quantum computation can be directly related to the presence of contextuality~\cite{Anders2009,RaussendorfContextuality,Raussendorf2023}.

Unfortunately, the only games known to provide quantum advantage throughout a phase employ symmetry-breaking phases~\cite{BulchandaniGames}, or, alternatively, symmetry-protected phases \cite{DanielStringOrder, Daniel2022Exp}, which are not robust to symmetry-breaking noise. This is also the only setting in which it is known how to tune across a phase transition involving a long-range-entangled phase using a finite-depth quantum circuit on available hardware \cite{chen2023realizing}.
\emph{Prima facie}, one might have thought it would be better to employ topologically ordered phases, which are robust to arbitrary local perturbations. However, it is not known how to robustly harness the quantum correlations inherent therein to gain quantum advantage, and, relatedly, it is not known how to tune across a topological phase transition using a finite-depth circuit on available hardware. How to realize, harness, and continuously tune topologically ordered phases on available quantum devices therefore remains an important open problem. 

\begin{figure*}[t]
    \centering
    \includegraphics[width=\linewidth]{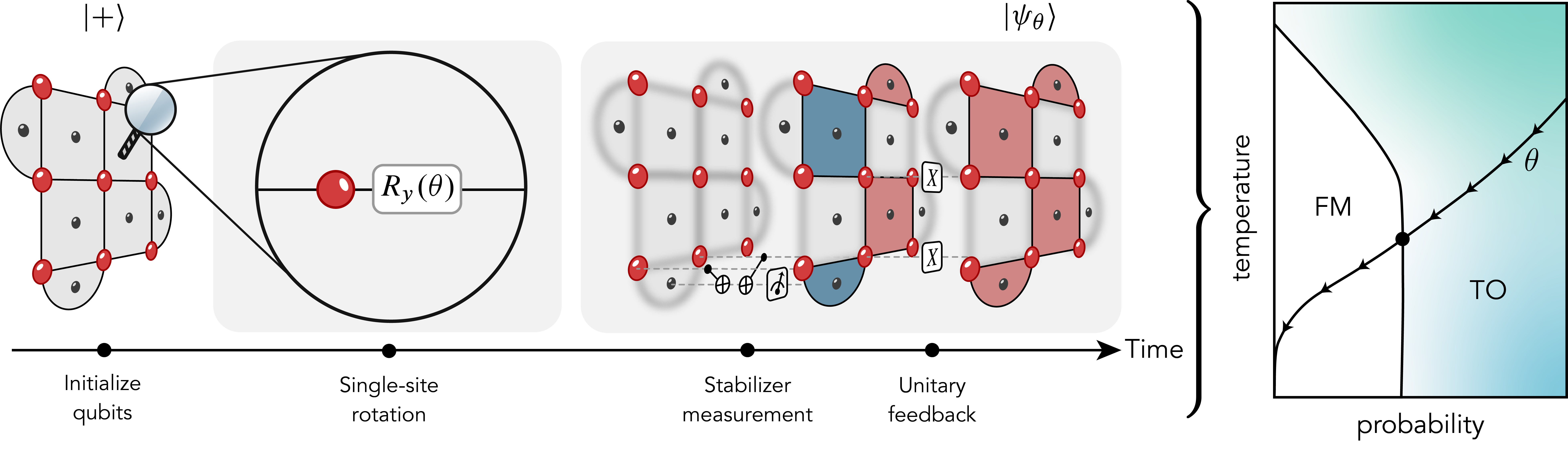}
    \caption{Schematic of the state preparation procedure. The data qubits (red circles) are initialized in the $\ket{+}$ state. Next, an $R_y(\theta)$ rotation is applied to every data qubit, which enacts the nonunitary operator~\eqref{eqn:nonunitary-Ry}. Next, RSC $Z$ stabilizers are measured by coupling to ancilla qubits (black circles), and $X$ operators are applied to correct any defective outcomes. Since the $X$ corrections do not commute with the $R_y(\theta)$ rotation, we arrive at a randomly deformed logical $\ket{\overline{+}}$ state of the form~\eqref{eqn:post-meas-TC}. Varying $\theta$ effectively traverses the phase diagram of the 2D RBIM~\eqref{eqn:pq-disorder-average}, shown in the $p$-$T$ plane, where $p$ is the probability of an antiferromagnetic bond. The paramagnetic phase is dual to a phase with topological order (TO).}
    \label{fig:cover-im}
\end{figure*}

In this Letter we introduce a family of multiplayer quantum games that harness topological order to gain robust quantum advantage, and demonstrate this advantage in experiments carried out on \emph{Quantinuum}'s H1-1 device. We also explain how one may robustly tune through a topological phase transition using a constant-depth circuit that utilizes mid-circuit measurement and feedback, which is witnessed by the presence of quantum advantage, or lack thereof. We present data on quantum hardware contrasting the robustness of quantum advantage secured by topologically ordered resource states with the more delicate advantage afforded by symmetry-broken states.

\textit{Parity games.}---The basic multiplayer game that we use is built upon the \emph{parity game} \cite{ghz1989,GHSZ1990,brassard2005pseudotelepathy} with $P \geq 3$ players who are not allowed to communicate during the game. The players are handed bits $x_i$, where $i=1, \dots, P$, such that $\sum_{i=1}^P x_i \equiv 0 \mod 2$. The players' task is to output bits $y_i$ such that
\begin{equation} \label{eq:winning_cond}
    \sum_{i=1}^P y_i \equiv \frac12 \sum_{i=1}^P x_i \mod 2\, .
\end{equation} 
We remark that the right-hand side can be equivalently written as $x_1\vee x_2\vee x_3$ where $\vee$ denotes the boolean `or' function, which connects the parity game to the use of quantum resources for computing nonlinear boolean functions~\cite{Anders2009,RaussendorfContextuality}. The optimal classical strategy~\cite{brassard2005recasting} wins the game with probability
\begin{equation}
    p_\text{cl} = \frac{1}{2} + \frac{1}{2^{\lceil P/2 \rceil}} \label{classicalbound}
    \, ,
\end{equation}
where $\lceil \, \cdot \, \rceil$ denotes the ceiling function. Simple deterministic classical strategies that optimize this bound are given in Ref.~\cite{brassard2005recasting}. However, it is known that this game can be won with certainty if the players share beforehand a $P$-qubit GHZ~\cite{ghz1989} state $|\text{GHZ}\rangle =  (|000\cdots\rangle + |111\cdots\rangle )/\sqrt{2}$. They simply measure their qubit in the $X$ ($Y$) basis if they receive $x_i=0$ ($x_i=1$), and output their measurement outcome. This may readily be verified to win the game with probability one. If the resource state is deformed from $|\text{GHZ}\rangle$ to $\ket{\psi}$ then the probability of victory becomes
\begin{equation}
    p_\text{q}(\ket{\psi}) = \frac12 \left[ 1 + \frac{1}{2^{P-1}} \langle \psi | M_P | \psi \rangle \right] \label{quantumvictory}
    \, ,
\end{equation}
where $M_P$ is a Mermin polynomial~\cite{MerminPolynomials}, which contains monomials of order $P$ composed of $X$ and $Y$ operators. For $P=3$ players, we have
\begin{equation}
    M_3 = X_1 X_2 X_3 - X_1 Y_2 Y_3 - Y_1 X_2 Y_3 - Y_1 Y_2 X_3
    \, .
    \label{eqn:M3}
\end{equation}
Quantum advantage exists IFF the victory probability \eqref{quantumvictory} exceeds the classical bound \eqref{classicalbound}. However, the quantum advantage in~\eqref{quantumvictory} is particularly sensitive to symmetry-breaking perturbations, which rapidly collapse the cat-like superposition to a product state. Indeed, in the large-$P$ limit, arbitrarily weak symmetry-breaking perturbations remove quantum advantage.% to disappear.

It is natural to ask whether topological order, which is known to be robust to arbitrary local perturbations, might afford more robust quantum advantage. However, the only known implementation of multiplayer games using topologically ordered resources~\cite{BBSgame} has no robustness in the thermodynamic limit and, moreover, requires the application of operators that can only be implemented with linear-depth circuits~\footnote{Specifically, one must apply ``square root of noncontractible Wilson loop'' operators, which cannot be locally decomposed.}. Here, we present a different family of multiplayer games, which realize the full potential of topological order by providing access to robust quantum advantage, and can be readily implemented on NISQ devices in finite depth. These games proceed by embedding the GHZ game described above into a topologically ordered phase. For specificity, we illustrate and implement the game with $P=3$ before generalizing to arbitrary $P$. We assume throughout that the resource state involves qubits arranged on the edges of a square lattice (which we often omit from our graphical notation, to avoid clutter).

\begin{figure*}[t]
    \centering
    \includegraphics[width=\linewidth]{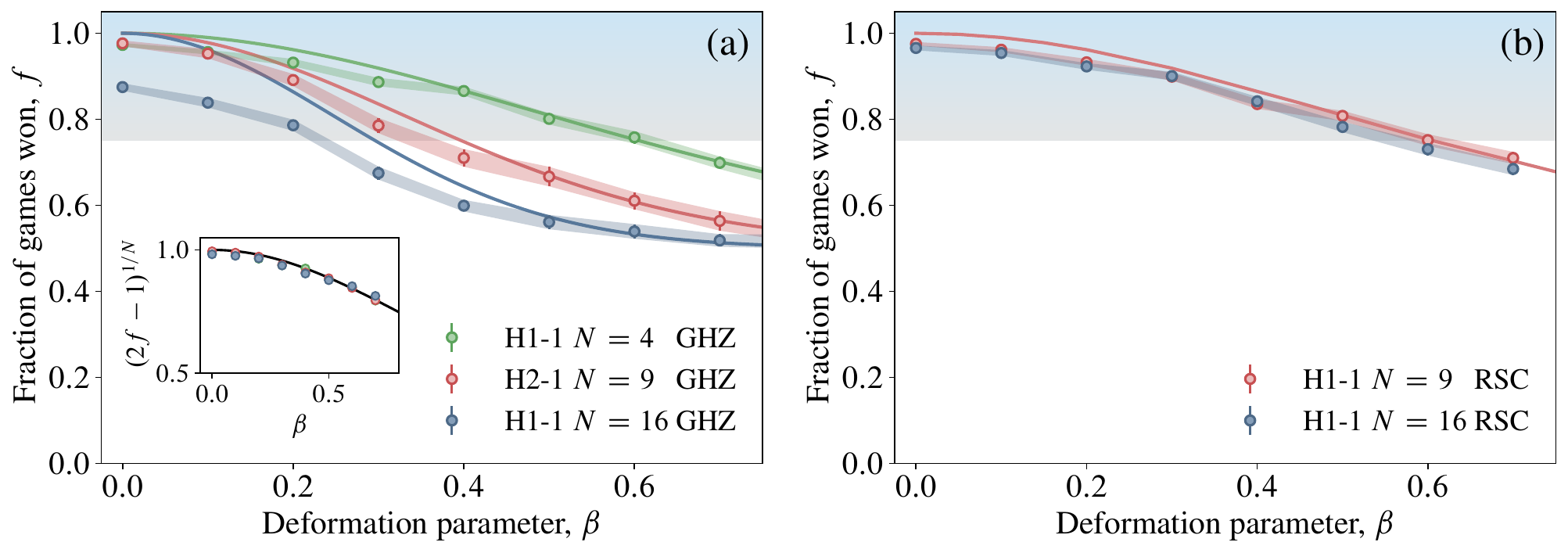}
    \caption{(a) Playing the parity game on Quantinuum's H1-1 and H2-1 quantum computers with GHZ states of varying size. We play the $P=3$ parity game with GHZ states of size $N=4, 9, 16$ qubits. The solid lines represent the exact expression for the victory probability~\eqref{eqn:GHZ-pq-exact}, while the shaded area represents the region of quantum advantage. In the inset, we illustrate that no robust quantum advantage exists in the thermodynamic limit with a data collapse. (b) Playing the game using the deformed $3\times3$ and $4\times4$ rotated surface codes. Unlike the GHZ state, quantum advantage remains robust as $N$ is increased. The data include one-$\sigma$ bootstrapped error bars. The solid lines (overlapping) correspond to the exact prediction based on exact numerical evaluation of Eq.~\eqref{eqn:pq-disorder-average}. All data points involve either 256 (H1-1) or 128 (H2-1) shots for each of the four unique inputs to the $P=3$ parity game.}
    \label{fig:H1-data}
\end{figure*}

\begin{figure}[t]
    \centering
    \includegraphics[width=\linewidth]{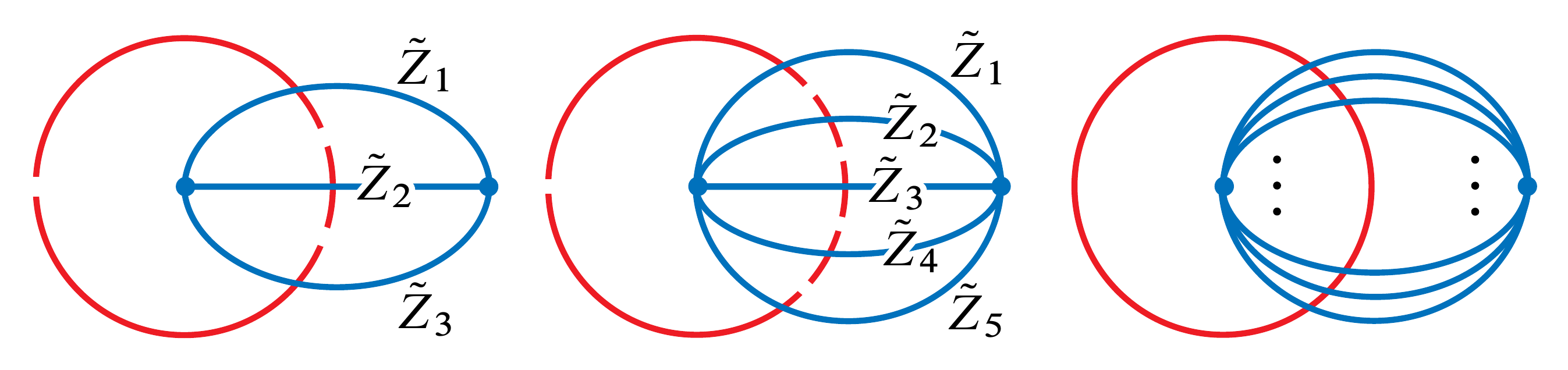}
    \caption{Illustration of the operator configurations that can be used to win the $P$-player parity game for $P=3, 5$, and generic $P$ from left to right. In all cases, the $X$ loop segments are implicitly labeled such that $\tilde{X}_i$ only anticommutes with $\tilde{Z}_i$.}
    \label{fig:p_game}
\end{figure}

{\it The $P=3$ game}.---We begin by assuming that the players share a ground state of the two-dimensional toric code~\cite{kitaev} at its solvable point, and by defining the operators
\begin{equation}
\arraycolsep=4pt\def\arraystretch{4}
\begin{array}{cccc}
    \tilde{X}_1 = \begin{tikzpicture}\Xbackground{0}{0};\Xtop{0}{0};\end{tikzpicture}
    &
    \tilde{X}_2 = \begin{tikzpicture}\Xbackground{0}{0};\Xsides{0}{0};\end{tikzpicture}
    &
    \tilde{X}_3 = \begin{tikzpicture}\Xbackground{0}{0};\Xbot{0}{0};\end{tikzpicture} 
    & {\displaystyle \tilde{X}_i = \prod_{e \in \ell^x_i} X_e} \\
    \tilde{Z}_1 = \begin{tikzpicture}\Zbackground{0}{0};\Ztop{0}{1};\end{tikzpicture}
    &
    \tilde{Z}_2 = \begin{tikzpicture}\Zbackground{0}{0};\ZH{0}{0};\end{tikzpicture} 
    &
    \tilde{Z}_3 = \begin{tikzpicture}\Zbackground{0}{0};\Zbot{0}{-1};\end{tikzpicture}
    & {\displaystyle \tilde{Z}_i = \prod_{e \in \ell^z_i} Z_e }
    \end{array}
    \label{eqn:TC-state-dependent-loops}
\end{equation}
Each red edge corresponds to the application of a Pauli $X$ operator on the direct lattice (light gray), and each blue line on the dual lattice corresponds to the application of a Pauli $Z$ operator [see~\eqref{eqn:mutual-stats} for the relative position of $\tilde{X}_i$ and $\tilde{Z}_i$ operators]. When acting on ideal toric code ground states, we have
\begin{subequations}
\begin{align}
    \tilde{X}_1 \tilde{Y}_2 \tilde{Y}_3 = \begin{tikzpicture}\XR{-0.5}{0.5};\ZL{0.5}{-0.5};\XL{-0.5}{0.5};\ZR{0.5}{-0.5};\end{tikzpicture}=-1 \quad \tilde{Y}_1 \tilde{Y}_2 \tilde{X}_3 = \begin{tikzpicture}\Xbot{-0.5}{-0.5};\Xsides{-0.5}{-0.5};\ZH{0.5}{-0.5};\Ztop{0.5}{0.5};\Xtop{-0.5}{-0.5};\end{tikzpicture}=-1 \\
    \tilde{Y}_1 \tilde{X}_2 \tilde{Y}_3 = \begin{tikzpicture}\Xbot{0}{0};\Zbot{1}{-1};\Xsides{0}{0};\Ztop{1}{1};\Xtop{0}{0};\end{tikzpicture}=-1 \quad
    \tilde{X}_1 \tilde{X}_2 \tilde{X}_3 = \begin{tikzpicture}\draw[white](0,0)to(2,0);\faceop{0}{0};\end{tikzpicture}=1
\end{align}%
\label{eqn:mutual-stats}%
\end{subequations}
which follows from fact that these operators can be interpreted as braiding diagrams, such that minus signs come from the mutual statistics of the anyonic excitations in the toric code. The operators in~\eqref{eqn:TC-state-dependent-loops} correspond to the smallest loops on which the game can be played. However, loops of any size and/or shape can be used, as long as they are homotopically equivalent to \eqref{eqn:mutual-stats}, see Fig.~\ref{fig:p_game}. 

To construct a game for which toric code ground states act as a resource, using the same nomenclature as Ref.~\cite{BBSgame}, associate a \emph{player} to each qubit belonging to the support of the loops in~\eqref{eqn:TC-state-dependent-loops}. The players belonging to the union of the support of $\tilde{X}_i$ and $\tilde{Z}_i$ will collectively be referred to as the $i$th \emph{team}. Each team has three types of player:
\begin{enumerate*}[label=(\roman*)]
    \item a \emph{mixed player}, who occupies the edge $e_i = \ell_i^x \cap \ell^z_i$, on which $\tilde{X}_i$ and $\tilde{Z}_i$ intersect,
    \item $X$ \emph{players}, occupying edges $\ell_i^x \setminus e_i$,
    \item $Z$ \emph{players}, occupying edges $\ell_i^z \setminus e_i$.
\end{enumerate*}
The teams are handed bits $x_i$ satisfying $\sum_{i=1}^3 x_i \equiv 0 \mod 2$, and players that belong to different teams may not communicate with one another. As in the parity game, the winning condition is that the output bits $y_i$ satisfy \eqref{eq:winning_cond}, which again has an optimal classical victory probability of $p_\text{cl}=3/4$~\eqref{classicalbound}. On the other hand, there exists a perfect strategy when the players share a toric code ground state (at the solvable point). If the $i$th team receives $x_i = 0$, the mixed player measures in the $X$ basis and the $X$ players measure their qubits in the $X$ basis.
If, however, the $i$th team receives $x_i = 1$, the mixed player measures in the $Y$ basis and the $X$ ($Z$) players measure their qubits in the $X$ ($Z$) basis. In either case, the team returns the output $y_i=\sum_k \lambda_k \mod 2$ where the sum runs over each measurement performed and $\lambda_k=0,1$ corresponds to measuring the eigenvalue $(-1)^{\lambda_k}$. Since this strategy is effectively measuring the operators in~\eqref{eqn:mutual-stats}, the players win with certainty if the resource state is a fixed-point toric code ground state. Furthermore, the success probability of the above protocol when armed with a generic resource state $\ket{\psi}$ is given by~\eqref{quantumvictory}, where the third-order Mermin polynomial is understood to contain the partial string operators in~\eqref{eqn:TC-state-dependent-loops}. Quantum advantage is lost when $\langle \tilde{M}_3 \rangle \le 2$.

We emphasize that the game has been constructed in such a way as to probe only \emph{local} data. Thus, the game, unlike the protocol in Ref.~\cite{BBSgame}, is not sensitive to boundary conditions, as long as the operators belong to the bulk. This reliance on purely local (but topological) data (the braiding statistics) also makes our game {\it robust to perturbations}, as we now discuss.

\textit{Deformed toric code states.---}%
To study the robustness of the game, we identify a family of states that 
\begin{enumerate*}[label=(\roman*)]
    \item can be created in constant depth using measurements and outcome-conditioned unitary feedback, and
    \item cross a topological phase transition as a function of a continuously tunable parameter that enters the circuit.
\end{enumerate*}
Consider the following family of Hamiltonians parametrized by the real parameter $\beta$
\begin{equation}
    H_\beta(\vec{s}) = - \sum_{v} A_v -  \sum_p B_p + \sum_p \exp\left(-\beta \sum_{e \in \partial p} s_e Z_e\right)
    \, .
    \label{eqn:deformed-random-TC}
\end{equation}
where $A_v = \prod_{e \in \partial^\dagger v} Z_e$ and $B_p  = \prod_{e \in \partial p} X_e$ are the standard stabilizer operators of Kitaev's toric code~\cite{Kitaev2003}, with $v$ and $p$ denoting vertices and plaquettes of the square lattice, respectively. The signs $s_e \in \{ +1, -1 \}$ are independent, identically distributed random variables. For all real $\beta$ and all sign configurations $\vec{s} = \{s_e\}$ on edges, it can be shown \cite{Castelnovo2008,Tsomokos2011} that 
\begin{equation}
    \ket{\psi_\beta(\vec{s})} = \frac{1}{\sqrt{Z}}\sum_{g \in G} \exp\left[\frac12 \beta \sum_{e} s_e Z_e(g)\right] g \ket{0}
    \, ,
    \label{eqn:post-meas-TC}
\end{equation}
is an exact ground state of~\eqref{eqn:deformed-random-TC}, where $G$ is the Abelian group generated by $B_p$ stabilizers, and $Z_e(g)$ is the $z$ component of the spin on edge $e$ in the product state $g\ket{0}$. This state can equivalently be obtained up to normalization by applying the operator $e^{\beta s_eZ_e/2}$ to every edge of a clean toric code ground state (ground state of $H_\beta(\vec{s})$ with $\beta=0$). The ability to write down such exact ground states is a general feature of generalized Rokhsar-Kivelson systems~\cite{Henley2004classical,Ardonne2004}, whose Hamiltonians obey a stochastic matrix form decomposition~\cite{Castelnovo2005}. Introducing Ising spin degrees of freedom $\theta_p$ living on the plaquettes, the loops on the direct lattice in $g\ket{0}$ map to domain walls between plaquette spins~\footnote{Note that the mapping from spin configurations on edges to Ising spins is two-to-one, with both $\{\theta_p\}$ and $\{-\theta_p\}$ representing the same domain-wall configuration (a global $\mathbb{Z}_2$ gauge redundancy).} (see also Refs.~\cite{Dennis2002,Wang2003}). Via this statistical mechanics mapping, we find that $\langle A_v \rangle = 1$, but $\langle B_p \rangle$ is degraded from its $\beta=0$ value, leading to a modified disorder-averaged victory probability 
\begin{equation}
    p_\text{q}(\beta) =\frac12 +  \mathbb{E}_{\mathbf{s}}  \left\langle \frac12\exp\left( -\beta \sum_{\langle p p' \rangle \in \partial p} \theta_p s_{pp'} \theta_{p'} \right) \right\rangle_\beta 
    \, ,
    \label{eqn:pq-disorder-average}
\end{equation}
where $\langle O(\vec{s}) \rangle_\beta$ is a classical thermal average of the disorder-dependent classical observable $O(\vec{s})$ with respect to the 2D random-bond Ising model (RBIM) Hamiltonian $H(\vec{s}) = -\sum_{\langle pp' \rangle} \theta_p s_{pp'} \theta_{p'}$ at inverse temperature $\beta$. The expectation value $\mathbb{E}_\vec{s}$ is a quenched disorder average over the variables $\vec{s}$ on edges. The paramagnetic and ferromagnetic phases of the classical Ising model map to the topological and trivial phases of~\eqref{eqn:deformed-random-TC}, respectively~\cite{Tsomokos2011}.
Monte Carlo simulations of the RBIM suggest that quantum advantage persists up to a deformation parameter $\beta \approx 0.6$ in the thermodynamic limit.

\textit{Device implementation and experimental results}.---%
We now describe the circuit that probabilistically creates states of the form~\eqref{eqn:post-meas-TC} in finite depth. Rather than working with deformed {toric code} ground states, we work instead with the \emph{rotated surface code} (RSC), and prepare the logical $\ket{\overline{+}}$ state thereof. The clean surface code ground state can be prepared using the measurement and feedback protocol outlined in Ref.~\cite{fossfeig2023experimental}, in which the RSC $Z$-type stabilizers are measured, and $X$-like strings are applied to correct defective measurement outcomes~\cite{Raussendorf2005LongRange}. The GHZ state can be created similarly \cite{Zhu2022}. To create the deformed state~\eqref{eqn:post-meas-TC}, we need to implement the nonunitary operator $e^{\pm\beta Z/2}$. One method to accomplish involves weakly coupling to an ancilla and measuring, i.e., a weak measurement~\cite{Zhu2022,chen2023realizing,LinImaginaryTime2021}. Here, we use an alternative method which involves no ancillas and only unitary operations. Namely, we simply act with the unitary rotation operator $R_y(\theta) = e^{i\theta Y/2}$ on $\ket{+}$ at the beginning of the state preparation procedure. This effectively enacts the required nonunitary deformation since
\begin{equation}
    e^{i\theta Y / 2}
    \ket{+} = (\cos\theta)^{1/2} e^{\beta Z / 2} \ket{+} 
    \, ,
    \label{eqn:nonunitary-Ry}
\end{equation}
if $\theta$ is chosen such that $\tan(\theta/2) = \tanh(\beta/2)$, with $\abs{\theta} < \pi/4$. Subsequent measurements of the $Z$-type stabilizers commute with this nonunitary deformation, but the $X$-like strings used to correct defective outcomes randomly flip the sign of $\beta$, resulting in states of the form~\eqref{eqn:post-meas-TC}, where the random signs $\vec{s}$ derive from the measurement outcomes for the stabilizers. Analogously to Refs.~\cite{Zhu2022,lee2022decoding,chen2023realizing}, the probability $p(\vec{s})$ of obtaining the sign configuration $\vec{s}$ follows the Nishimori line~\cite{Nishimori_I, Nishimori_II} of the random-bond Ising model~\footnote{Technically, only operators that are invariant under the choice of $X$-like correction procedure (``gauge-invariant operators'') can be mapped to thermal averages along the Nishimori line.}, i.e., $p(s) = (1+e^{-2s\beta})^{-1}$ on each edge (shown schematically in Fig.~\ref{fig:cover-im}). We can similarly prepare a deformed GHZ state with $e^{\pm\beta Z/2}$ (which plays the role of a symmetry-breaking deformation) applied to each site.

The result of playing the games on the Quantinuum H1-1 quantum computer using this preparation procedure is shown in Fig.~\ref{fig:H1-data}. When using $N$-qubit GHZ states as resources, it is necessary to measure $N-3$ of the qubits in the $X$ basis and play the parity game with the remaining 3 qubits \cite{BulchandaniGames}.  
In Fig.~\ref{fig:H1-data}(a) we track the degradation of quantum advantage as a function of symmetry-breaking deformation strength; quantum advantage is lost more rapidly as the size of the GHZ state is increased, as verified by the collapse of the data in the inset. Note that the symmetry-breaking deformation leads to advantage over a window with central limit theorem scaling, $N^{-1/2}$ for an $N$-qubit GHZ state, since
\begin{equation}
    p_\text{q}(\beta) = \frac12 [1 + \sech(\beta)^N]
    \, .
    \label{eqn:GHZ-pq-exact}
\end{equation}
A local Hamiltonian perturbation (e.g., a longitudinal field) would be even more detrimental, producing an exponentially small region of advantage in $N$.
In contrast, when using the RSC as a resource, we only need to measure the qubits that contribute to the operators in \eqref{eqn:mutual-stats}. Thanks to this, the toric code game introduced herein exhibits essentially no dependence on system size [Fig.~\ref{fig:H1-data}(b)]; there is robust quantum advantage through a large swath of the topologically ordered phase.

{\it Increasing the number of players}.---%
We now discuss how toric code ground states may be used to win generalized games with $P \geq 3$ players.
The games have the same inputs and outputs as the $P$-player parity game, so we need only discuss the relevant operators to be measured. The toric code state will continue to serve as a resource conferring quantum advantage, since it has $P$-partite entanglement for any $P$. The motivation for using toric code states over GHZ states to win $P$-player games is the following. As $P$ is increased, it is necessary to create larger resource states, which amplifies the effects of local errors or perturbations. This results in a critical value of $P$ beyond which the quantum advantage in the $P$-player game is lost. Since large GHZ states are less robust than large toric code states, as demonstrated above, one naturally expects that using toric codes states allows us to reach much larger values of $P$ before losing quantum advantage.

The strategy for the $P=3$-player game can be summarized by the leftmost loop diagram in Fig.~\ref{fig:p_game}. Generalizing this diagram to arbitrary $P$ defines the operators $\tilde{X}_i$ and $\tilde{Z}_i$ for $i=1, \dots, P$, which satisfy $\tilde{X}_i \tilde{Z}_j = (-1)^{\delta_{ij}}\tilde{Z}_j\tilde{X}_i$ and, when acting on ideal toric code ground states, $\prod_i \tilde{X}_i=\tilde{Z}_j\tilde{Z}_k=1$, see Fig.~\ref{fig:p_game}. This allows us to directly employ the strategy that was used to win the $P$-player GHZ game. That is, the $i$th team measures $\tilde{X}_i$ ($\tilde{Y}_i$) when they receive $x_i=0$ ($x_i=1$). This naturally produces the correct outputs as a result of the braiding statistics of anyons. For instance, when $P=5$,
\begin{equation}
    \tilde{Y}_1 \tilde{Y}_2 \tilde{X}_3 \tilde{Y}_4 \tilde{Y}_5 = 
    \begin{tikzpicture}
        \draw [xpauli, line width=1.25] (2, -1) -- (2, -1) -- (2, -1) -- (8, -1);
        \draw [xpauli, line width=1.25] (2, 1) -- (2, 1) -- (2, 1) -- (8, 1);
        \begin{scope}[very thick,decoration={markings,
            mark=at position 0.125 with {\fill[white] +(-2pt,-2pt) rectangle +(2pt,2pt);},
            mark=at position 0.375 with {\fill[white] +(-2pt,-2pt) rectangle +(2pt,2pt);},
            mark=at position 0.625 with {\fill[white] +(-2pt,-2pt) rectangle +(2pt,2pt);}}
            ]
            \draw [line width=1.25, zpauli, preaction={decorate}] (1, 0) rectangle (3, 2);
            \draw [line width=1.25, zpauli, preaction={decorate}] (7, 0) rectangle (9, 2);
        \end{scope}
        \begin{scope}[very thick,decoration={markings,
            mark=at position 0.25 with {\fill[white] +(-2pt,-2pt) rectangle +(2pt,2pt);}}
            ]
            \draw [line width=1.25, xpauli, preaction={decorate}] (2, -1) rectangle (2, 1);
            \draw [line width=1.25, xpauli, preaction={decorate}] (8, -1) rectangle (8, 1);
        \end{scope}
    \end{tikzpicture}
    =+1
    \, .
\end{equation}

\textit{Discussion.---}%
We have demonstrated how topological order may be harnessed to gain robust quantum advantage at the task of winning a nonlocal game. This may enable novel routes to secure nonlocal coordination without classical communication. We have also shown how one may continuously tune across a topological phase transition with a finite-depth quantum circuit by harnessing measurements and unitary feedback. Generalizations of the results presented herein, as well as the broader connections to quantum order and disorder operators, contextuality, and the homology of loop braiding, will be presented elsewhere~\cite{inprep}. 

\begin{acknowledgments}
We thank Karl Mayer for helpful discussions during the early stages of this project. RN and OH are supported by the U.S.~Department of Energy, Office of Science, Basic Energy Sciences, under Award DE-SC0021346. This research used resources of the Oak Ridge Leadership Computing Facility, which is a DOE Office of Science User Facility supported under Contract DE-AC05-00OR22725. DTS is supported by the Simons Collaboration on Ultra-Quantum Matter, which is a grant from the Simons Foundation (651440).
\end{acknowledgments}

%%%%%%%%%%%%%%%%%%%%%%%%%%%%%%%%%%%%%%%%%%%%%%%%%%%%%%%%%%%%%%%
%%%%%%%%%%%%%%%%%%%%%%%%%%%%%%%%%%%%%%%%%%%%%%%%%%%%%%%%%%%%%%%
%%%%%%%%%%%%%%%%%%%%%%%%%%%%%%%%%%%%%%%%%%%%%%%%%%%%%%%%%%%%%%%

\bibliography{biblio.bib}

\end{document}